\documentclass[twocolumn,showpacs,preprintnumbers,amsmath,amssymb]{revtex4}
\usepackage{graphicx}
\usepackage{dcolumn}
\usepackage{bm}
\usepackage{amssymb}
\usepackage{graphicx}
\usepackage{amsmath}
\usepackage{xspace}

\begin{document}

\title{Raman scattering study of (K$_x$Sr$_{1-x}$)Fe$_2$As$_2$ ($x = 0.0, 0.4)$}
\author{A.~P.~Litvinchuk$^{1,2}$, V.~G.~Hadjiev$^{1,3}$, M. N. Iliev$^{1,2}$,
    Bing Lv$^{1,4}$, A.~M.~Guloy$^{1,4}$, and C.~W.~Chu$^{1,2,5}$}
\affiliation{
$^1$Texas Center for Superconductivity at the University of Houston, TX 77204-5002\\
$^2$Department of Physics, University of Houston, Texas 77204\\
$^3$Department of Mechanical Engineering, University of Houston, Texas 77204\\
$^4$Department of Chemistry,University of Houston, Texas 77204\\
$^5$Hong Kong University of Science and Technology, Hong Kong,
China }
\date{\today}

\begin{abstract}
Polarized Raman spectra of non-superconducting SrFe$_2$As$_2$ and
superconducting K$_{0.4}$Sr$_{0.6}$Fe$_2$As$_2$ ($T_c = 37$~K) are
reported. All four phonon modes (A$_{1g}$ + B$_{1g}$ + 2E$_g$)
allowed by symmetry, are found and identified. Shell model gives
reasonable description of the spectra. No detectable anomalies are
observed near the tetragonal-to-orthorhombic transition in
SrFe$_2$As$_2$ or the superconducting transition in
K$_{0.4}$Sr$_{0.6}$Fe$_2$As$_2$.
\end{abstract}

\pacs{74.72.-b, 74.25.Kc, 63.20.D-, 78.30.-j}
\maketitle

The renewed interested in superconductors was sparked recently by
the discovery of a novel class of iron-arsenide based oxypnictides
RFeAsO$_{1-x}$F$_x$ (where R is a rare earth
element).\cite{ROFA1,ROFA2,ROFA3,ROFA4,ROFA5} Similarly to the
cuprate superconductors, doping of superconducting FeAs planes in
oxypnictides determines one of the key characteristics of a given
material, its superconducting transition temperature, which
reaches values as high as T$_c$=54~K. Oxypnictides were shown to
exhibit {\it n}-type conductivity. More recently a series of
compounds with similarly structured FeAs planes, but different
charge reservoir block, A$_x$M$_{1-x}$Fe$_2$As$_2$ (where A is an
alkali element and M - Sr or Ba) were found to exhibit
superconducting properties.\cite{rotter} Unlike oxypnictides these
new superconductors clearly show {\it p}-type
conductivity.\cite{KSFA1} Optimal material doping is achieved in
this later system at $x \approx 0.4-0.5$ when the critical
temperature reaches T$_c \sim 38$~K.\cite{KSFA2}

First principle electronic band structure calculations of
oxypnictides point toward unconventional superconductivity, that
is mediated by antiferromagnetic spin
fluctuations\cite{calc1,calc2,calc3}. Due to rather weak
electron-phonon interactions it is generally believed that the
phonons do not contribute substantially to the superconductivity.
Despite this fact, it is important to experimentally identify the
symmetry and frequency of phonon excitations and search for
specific features in the Raman scattering spectra, which could
shed light onto the properties of the superconducting state. In
this communication we report the results of polarized Raman
scattering studies of (K$_x$Sr$_{1-x}$)Fe$_2$As$_2$ for the parent
$x=0$ compound and superconducting material with $x=0.4$. We also
present the results of shell model lattice dynamics calculations,
which are in a good agreement with the experimental data.

The compounds under investigation were prepared by
high-temperature solid state reactions of high purity K and Sr
with FeAs, as described elsewhere.\cite{KSFA2} For the mixed-metal
samples, (K$_x$Sr$_{1-x}$)Fe$_2$As$_2$, stoichiometric amounts of
the ternary iron arsenides were thoroughly mixed, pressed and then
annealed within welded Nb containers (jacketed in quartz) at about
900~C for 70-120 hours. The polycrystalline samples containing
microcrystals of typical size $50\times 50\times 2$~$\mu$m$^3$
were characterized by X-ray diffraction, resistivity, magnetic
susceptibility, Hall and thermoelectric power
measurements.\cite{KSFA2,lorenz} Raman scattering measurements
were performed with a triple Horiba~JY T64000 spectrometer,
equipped with an optical microscope and liquid-nitrogen-cooled CCD
detector. Samples were mounted on the cold finger of an optical
cryostat. He-Ne laser ($\lambda_{las}=638.2$nm) was used for the
excitation and the power density did not exceed $10^4$~W/cm$^2$ in
order to minimize heating of the sample. Similarly to the case of
RFeAsO\cite{hadjiev08} the Raman intensities were very low and
required long acquisition time.

AFe$_2$As$_2$ crystallizes in the tetragonal ThCr$_2$Si$_2$ type
structure with space group I4/mmm (D$_{4h}^{17}$).\cite{rozsa} The
unit parameters and atomic positions were obtained from powder
diffraction data, using Rietveld refinement. The atomic positions
were consistent with the literature values. The cell parameters
for SrFe$_2$As$_2$ are $a = 0.39259(2)$~nm, $c = 1.2375(1)$~nm,
for K$_0.40$Sr$_0.60$Fe$_2$As$_2$ are $a = 0.38898(2)$~nm, $c =
1.2948(1)$~nm;  $z_{As} = 0.3516$ for both compounds. The
structure features individual FeAs layers identical to those in
RFeAsO, but with a different layer stacking sequence (AA in
RFeAsO, and AB in the ThCr$_2$Si$_2$-type structure).

From symmetry considerations\cite{russo} one expects four
Raman-active phonons: A$_{1g}$(As), B$_{1g}$(Fe), E$_g$(As), and
E$_g$(Fe) (Table~I). Using the polarization selection rules and
the fact that the $ab$ surfaces could easily be visually
recognized, the identification of the Raman line symmetry is
straightforward. In particular, in the spectra obtained from the
$ab$ plane the intensity of the $A_{1g}$ mode will remain constant
for any orientation of the incident polarization
$\overrightarrow{e_i}$ given that the scattered polarization
$\overrightarrow{e_s}$ is parallel to it ($\overrightarrow{e_s}
\parallel \overrightarrow{e_i}$) and will be zero in any crossed
polarization configuration ($\overrightarrow{e_s} \perp
\overrightarrow{e_i}$). The intensity of the B$_{1g}$ mode,
however, depends on the angle $\alpha$ between
$\overrightarrow{e_i}$ and the $a$-axis being proportional to
${\cos}^2 2\alpha$ for parallel and ${\sin}^2 2\alpha$ for crossed
configuration. Figure~1 shows the variations with $\alpha$ of the
Raman spectra of SrFe$_2$As$_2$ obtained from the $ab$ plane. Only
one line at 204 cm$^{-1}$ is clearly pronounced in the spectra and
its intensity follows the angular dependence expected for the
B$_{1g}$ mode. Obviously, the intensity of the A$_{1g}$ mode is
negligible for incident light propagating along the $c$-axis and
polarized in the $ab$ plane. However, all Raman active phonons are
clearly seen in the spectra taken from $ac$($bc$) surface, as
illustrated in Fig.~2. These observations suggest the following
relations between the elements of the Raman tensor in Table~I:
$|b|\gg |a|$ and $|c| \gg |a|$. The experimental phonon
frequencies are listed in Table~I.

\begin{figure}[]
\includegraphics[width=7.5 cm]{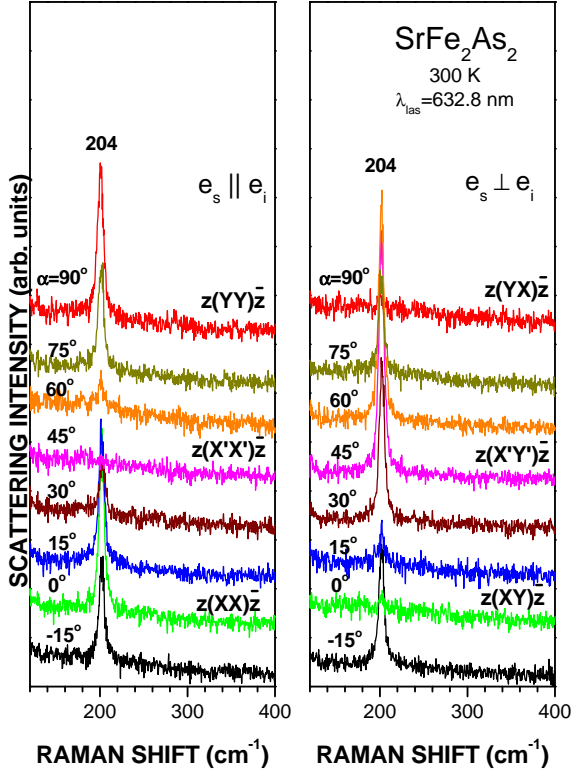}
\caption{(Color online) Raman spectra of SrFe$_2$As$_2$ for a
series of $ab$-plane polarized scattering geometries. Numbers on
the left denote the angle $\alpha$ between incident light
polarization and the crystallographic $a$-axis. }
\end{figure}
\begin{figure}
\includegraphics[width=7.5cm]{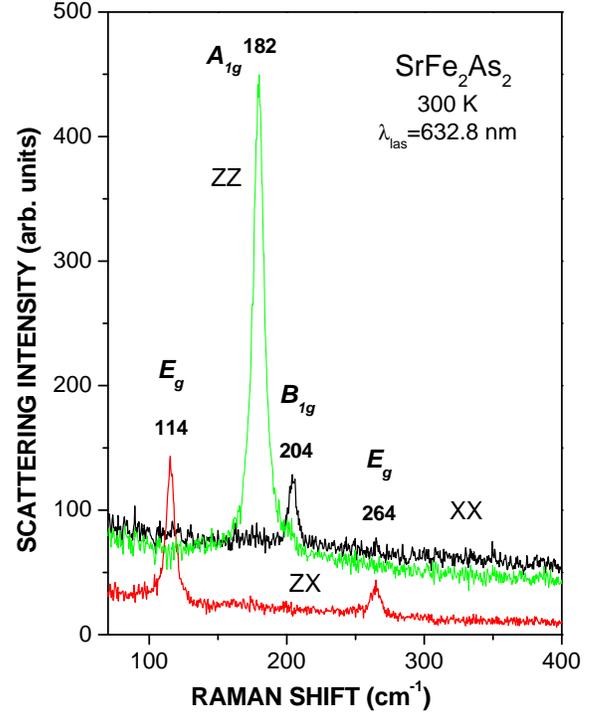}
\caption{$XX$ (Color online) A$_{1g}$+B$_{1g}$ modes allowed),
$ZX$ (E$_g$), and $ZZ$ (A$_{1g}$) spectra of SrFe$_2$As$_2$, as
obtained from $ac$ surface at room temperature.}
\end{figure}

\begin{table}[htb]
\caption []{Wyckoff positions and irreducible representations for
SrFe$_2$As$_2$ (space group $I4/mmm$, No.139), which yield
Brillouin zone center modes. Lower part of the Table lists
experimental and calculated mode frequencies and their activity.}
\begin{ruledtabular}
\begin{tabular}{c c c  }

    & Wyckoff         & $\Gamma$-point \\
Atom    & position     & phonon modes   \\
\hline
  &  &    \\
Sr  & 2a        & $         A_{2u} +                  E_u$  \\
Fe  & 4d        & $         A_{2u} + B_{1g} +   E_g + E_u$  \\
As  & 4e        & $A_{1g} + A_{2u}+             E_g + E_u$ \\
  &  &    \\
\multicolumn{3}{c}{Modes classification:} \\

\multicolumn{3}{c}{$\Gamma_{\rm Raman} = A_{1g} + B_{1g} + 2E_g$}\\
\multicolumn{3}{c}{$\Gamma_{\rm IR} = 2A_{2u} + 2E_u$}\\
\multicolumn{3}{c}{$\Gamma_{\rm Acoustic} = A_{2u} + E_u$}\\
  &  &  \\
\multicolumn{3}{c}{Raman tensors:} \\

\multicolumn{3}{c}{A$_{1g}(x^2+y^2,z^2) \rightarrow \left[
\begin{array}{ccc}
 a & 0 & 0 \\
 0 & a & 0 \\
 0 & 0 & b
 \end{array}
 \right]$} \\

 & \\

 \multicolumn{3}{c}{B$_{1g}(x^2-y^2) \rightarrow \left[
\begin{array}{ccc}
 c & 0 & 0 \\
 0 & -c & 0 \\
 0 & 0 & 0
 \end{array}  \right]$}\\

  & \\

 \multicolumn{3}{c}
 {$E_{g_1}(xz),E_{g_2}(yz) \rightarrow \left[
\begin{array}{ccc}
 0 & 0 & -e \\
 0 & 0 & 0 \\
 -e & 0 & 0
 \end{array}\right]$,
 $\left[ \begin{array}{ccc}
 0 & 0 & 0 \\
 0 & 0 & e \\
 0 & e & 0
 \end{array}\right]$ }  \\
    &  \\
\end{tabular}
\end{ruledtabular}

\begin{tabular}{|c|c|c|c|c|c|}
Mode  & Type & exp & LDC & Main atomic  & Allowed \\
      &      & cm$^{-1}$& cm$^{-1}$& displacements & polarizations\\
      \hline
A$_{1g}$ & Raman & 182 & 183 & As($z$)            & $XX$, $YY$, $ZZ$ \\
B$_{1g}$ & Raman & 204 & 203 & Fe($z$)            & $XX$, $YY$, $X^{\prime}Y^{\prime}$ \\
E$_{g}$  & Raman & 114 & 111 & As($xy$), Fe($xy$) & $XZ$, $YZ$ \\
E$_{g}$  & Raman & 264 & 335 & Fe($xy$), As($xy$) & $XZ$, $YZ$ \\
A$_{2u}$ & IR    &     & 198 & Sr($z$), As($-z$)  & Z \\
A$_{2u}$ & IR    &     & 322 & Fe($z$), Sr($-z$)  & Z \\
E$_{u}$  & IR    &     & 135 & Sr($xy$)           & X, Y \\
E$_{u}$  & IR    &     & 263 & Fe($xy$), As($-xy$)& X, Y \\
\hline
\end{tabular}
\end{table}

We also performed shell model calculations of the lattice dynamics
using the General Utility Lattice Program (GULP)\cite{gulp}, which
is known to reasonably describe wide class of ionic materials,
oxides in particular.\cite{popov,iliev} In the shell model each
ion is considered as a point core with charge $Y$ surrounded by a
massless shell with charge $Q$. The free ion polarizability is
accounted for by the force constant $k$. The short range
potentials $V(r)$ are chosen in the Born-Mayer-Buckingham form

$$ V(r) = a\exp(-r/r_{\rm 0}) - cr^{-6}.$$

The model parameters of Fe$^{2+}$, Sr$^{2+}$ and As$^{3-}$ were
tuned in order to achieve the best agreement with experimental
even-parity phonon modes (to the best of our knowledge, there is
no experimental information on odd-parity infrared-active
phonons); they are listed in Table~II. The displacement patterns
of all four Raman-active modes are shown in Fig.~3. It is worth
noting here that while the A$_{1g}$ and B$_{1g}$ modes can be
considered as "pure" modes, involving displacements along the
$c$-axis of either As or only Fe, the two E$_g$ modes are strongly
mixed. As expected, the A$_{1g}$ and B$_{1g}$ modes of
SrFe$_2$As$_2$ are very close in frequency to the corresponding
modes of same symmetry in RFeAsO.\cite{hadjiev08} Indeed, the
individual FeAs layers in both compounds are practically identical
in term of bond lengths and angles.

\begin{figure}[]
\includegraphics[width=5cm]{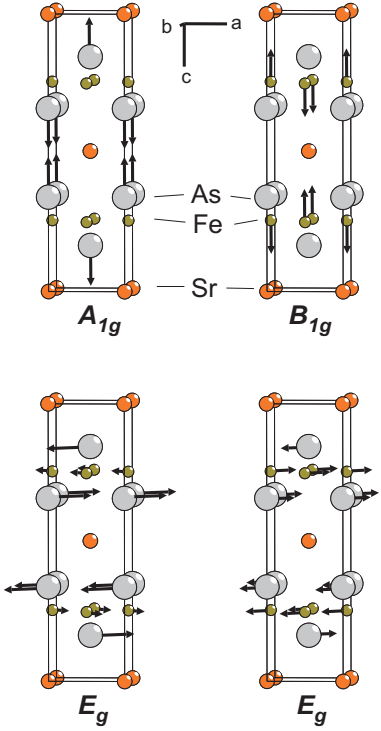}
\caption{(Color online) Displacement patterns of Raman-active
modes of SrFe$_2$As$_2$ from the shell model calculations. }
\end{figure}

\begin{table}
\caption []{Shell model parameters and the short range potentials
for SrFe$_2$As$_2$. We used structural parameters reported in
Ref.~\onlinecite{KSFA2} and listed in the text.}
\begin{tabular}{c c c c | c r c c}
\hline Ion &$Y(|e|)$&$Q(|e|)$&$k$(eV$\times$\AA$^{-2}$)&Ion
pair&$a$(eV)&$r_{\rm 0}$(\AA)&$c$(eV$\times$\AA$^6$)\\
\hline
Sr  & 2.40 & -0.50 &  10.0 &  Sr-As  &    1226 &  0.482  & 0   \\
Fe  & 2.40 & -0.50 &  19.9 &  Fe-As  &    2399 &  0.350  & 0   \\
As  & 0.15 & -3.00 &  15.2 &  As-As  &    2000 &  0.209  & 2500\\
\hline
\end{tabular}
\end{table}

\begin{figure}
\includegraphics[width=8cm]{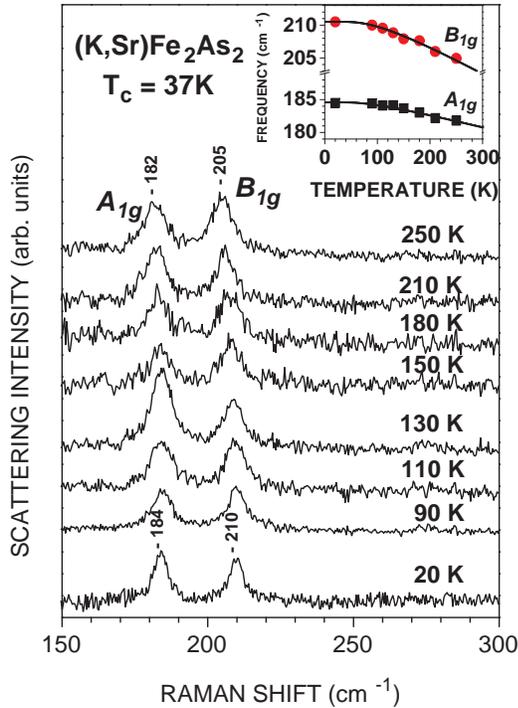}
\caption{(Color online) Temperature dependent Raman spectra of
K$_{0.4}$Sr$_{0.6}$Fe$_2$As$_2$.
}
\end{figure}

Due to the small size of crystallites and long acquisition time,
the Raman spectra of superconducting
K$_{0.4}$Sr$_{0.6}$Fe$_2$As$_2$ ($T_c = 37$~K) could not be
obtained in an exact scattering configuration and only the
$A_{1g}$ and $B_{1g}$ modes were observed. Their frequency is
practically the same as in the parent compound, the line width
being somewhat larger. Upon lowering temperature these two modes
show standard anharmonic behavior in both materials with rather
moderate frequency and width variations as illustrated in Fig.~4
for K$_{0.4}$Sr$_{0.6}$Fe$_2$As$_2$. No phonon anomalies either
near the tetragonal-to-orthorhombic structural transition of
SrFe$_2$As$_2$ at $T_{t-o} \approx 203$~K\cite{tegel}, or below
T$_c$=37~K of K$_{0.4}$Sr$_{0.6}$Fe$_2$As$_2$ were observed within
the accuracy of our experiments for the A$_{1g}$ and B$_{1g}$
modes. In principle, the main effect of the structural transition
should be a splitting of the E$_g$ modes, which position was
difficult to follow in the temperature-dependent experiments. The
reported anisotropy of Fe-As bond lengths within ab-plane,
however, is only 0.55\%\cite{tegel}. Following a simple
expression\cite{hadjiev08} for the phonon frequency $\omega_{ph}$
as a function of bond length $l$: $\omega_{ph}^2 \sim  1/l^3$, one
might expect splitting of the modes by about 0.8\%. Even for the
high frequency E$_g$ mode at 264 cm$^{-1}$ this yields 2.1
cm$^{-1}$, which is small compared to the linewidth and is
challenging  to be observed experimentally.

In conclusion, all four Raman active phonons in SrFe$_2$As$_2$
have been observed experimentally. The substitution of K for Sr
has little effect on the frequencies of Raman modes involving As
and Fe vibrations. The structural transition in SrFe$_2$As$_2$ and
superconducting transition in K$_{0.4}$Sr$_{0.6}$Fe$_2$As$_2$ do
not produce detectable anomalies in the parameters of A$_{1g}$ and
B$_{1g}$ modes.

\acknowledgements This work is supported in part by the T.L.L.
Temple Foundation, the J.J. and R. Moores Endowment, the State of
Texas through TCSUH, the USAF Office of Scientific Research, and
the LBNL through USDOE. B.L. and A.M.G. acknowledge the support
from the NSF (CHE-0616805) and the R.A. Welch Foundation. We are
grateful to Zhongjia Tang for help with crystallographic
calculations.

\end{document}